\documentclass[
superscriptaddress, 
preprint,
amsmath,amssymb, aps, prb
]{revtex4-2}

\usepackage{xcolor}
\usepackage{graphicx}% Include figure files
\usepackage{dcolumn}% Align table columns on decimal point
\usepackage{bm}% bold math
\usepackage[colorlinks,allcolors=blue,urlcolor=blue]{hyperref}
\usepackage[normalem]{ulem}

\usepackage{amsmath}
\usepackage{multirow} 
\usepackage{amssymb}
\usepackage[linesnumbered,ruled,vlined]{algorithm2e} 
\usepackage{listings}

\newcommand{\ket}[1]{\left| #1 \right>} % for average
\begin{document}

\title{Efficient GPU Parallelization of Electronic Transport and Nonequilibrium Dynamics from Electron-Phonon Interactions in the {\sc Perturbo} Code}

\author{Shiyu Peng}
\thanks{These authors contributed equally to this work.}
\affiliation{Department of Applied Physics and Materials Science, and Department of Physics, California Institute of Technology, Pasadena, California 91125, USA}

\author{Donnie Pinkston}
\thanks{These authors contributed equally to this work.}
\affiliation{Schmidt Academy for Software Engineering, California Institute of Technology, Pasadena, California 91125, USA}

\author{Jia Yao}
\affiliation{Department of Applied Physics and Materials Science, and Department of Physics, California Institute of Technology, Pasadena, California 91125, USA}

\author{Sergei Kliavinek}
\affiliation{Department of Applied Physics and Materials Science, and Department of Physics, California Institute of Technology, Pasadena, California 91125, USA}

\author{Ivan Maliyov}
\affiliation{Department of Applied Physics and Materials Science, and Department of Physics, California Institute of Technology, Pasadena, California 91125, USA}

\author{Marco Bernardi\thanks{}}\email{bmarco@caltech.edu}
\affiliation{Department of Applied Physics and Materials Science, and Department of Physics, California Institute of Technology, Pasadena, California 91125, USA}

% ABSTRACT
\maketitle
\noindent
The Boltzmann transport equation (BTE) with electron-phonon ($e$-ph) interactions computed from first principles is widely used to study electronic transport and nonequilibrium dynamics in materials. Calculating the $e$-ph collision integral is the most important step in the BTE, but it remains computationally costly, even with current MPI+OpenMP parallelization. 
This challenge makes it difficult to study materials with large unit cells and to achieve high resolution in momentum space.
Here, we show acceleration of BTE calculations of electronic transport and ultrafast dynamics using graphical processing units (GPUs). We implement a novel data structure and algorithm, optimized for GPU hardware and developed using OpenACC, to process scattering channels and efficiently compute the collision integral. This approach significantly reduces the overhead for data referencing, movement, and synchronization. Relative to the efficient CPU implementation in the open-source package \textsc{Perturbo} (v2.2.0), used as a baseline, this approach achieves a speed-up of $40$ times for both transport and nonequilibrium dynamics on GPU hardware, and achieves nearly linear scaling up to 100 GPUs. 
The novel data structure can be generalized to other electron interactions and scattering processes. We released this GPU implementation in the latest public version (v3.0.0) of \textsc{Perturbo}. The new MPI+OpenMP+GPU parallelization enables sweeping studies of $e$-ph physics and electron dynamics in conventional and quantum materials, and prepares \textsc{Perturbo} for exascale supercomputing platforms.\vspace{-18pt} 

\setcitestyle{super}

\newpage
\section{Introduction}
\vspace{-6pt}
Electron-phonon ($e$-ph) interactions govern a wide range of properties in materials, including electronic transport, nonequilibrium dynamics of excited electrons, superconductivity, optical behavior, and polarons~\cite{ziman2001electrons, mahan2011condensed}. 
First-principles calculations based on density functional theory (DFT) and related techniques can calculate $e$-ph interactions with quantitative accuracy~\cite{bernardi2016first, giustino2017electron}. 
In recent years, the combination of semiclassical Boltzmann transport equation (BTE) with first-principles $e$-ph interactions has enabled accurate predictions of electronic transport in metals~\cite{park2014electron, Mustafa}, inorganic and organic semiconductors~\cite{wuli2015, Zhou2016, Liu-first, naph, yliu, ponce2020first, Desai, chang2022, chang_bandlike_2025}, complex oxides~\cite{STO-tdep,zhou2019predicting,yao_DMC}, and quantum materials~\cite{David_prm, David_prl, Gao_tbg, Desai_2d}.
\\
\indent
For studies of ultrafast nonequilibrium dynamics, solving the BTE in the time domain $-$ a scheme called the real-time BTE (rt-BTE) $-$ provides a favorable balance between accuracy and computational cost~\cite{Bernardi_2023, Caruso-rev}. At present, this method can address the ultrafast dynamics of excited electrons~\cite{bernardi-si,Jhalani2017,sjakste2018hot}, the coupled nonequilibrium dynamics of electrons and phonons~\cite{tong2021toward, Caruso_2021, Caruso-PRX, kelly-ph}, and with appropriate extensions to the formalism, the ultrafast dynamics of excitons~\cite{chen2020exciton, Chen-excitons}. 
To accelerate the solution, the rt-BTE method can also take advantage of data-driven techniques~\cite{brunton2022data}, such as dynamic mode decomposition~\cite{maliyov2024dynamic, reeves2023dynamic} and interaction compression~\cite{luo2024data}, as well as advanced adaptive and multi-rate time-stepping schemes~\cite{kelly-ph}. 
The rt-BTE scheme offers a practical approach focused on incoherent dynamics on femtosecond to picosecond timescales that complements methods targeting coherent dynamics, such as time-dependent DFT~\cite{tancogne2020octopus, falke2014coherent} and nonequilibrium Green's functions~\cite{Sangalli2015, Perfetto2022}.
\\
\indent
Despite the overall efficiency, the rt-BTE method still requires parallelization and extensive software optimization, particularly for simulations of materials with large unit cells, and/or using dense momentum grids and targeting long simulation times beyond the picosecond timescale~\cite{kelly-ph}.
In the {\sc Perturbo} code, after identifying the relevant $e$-ph scattering processes, the collision integral $-$ the key quantity in BTE calculations $-$ is computed by looping over these scattering channels~\cite{zhou2021perturbo}. 
Even after selecting a relevant energy window and retaining only energy-conserving scattering channels, in a typical calculation, the total number of active scattering channels can still be as large as $10^8$ or higher, which poses a major computational challenge for CPU hardware. 
Therefore, for both transport and time-domain dynamics, it is particularly important to design a data structure that addresses the high-dimensionality and sparsity of $e$-ph interactions and scattering processes~\cite{kelly-ph}.
\\
\indent
Using graphic processing units (GPUs)~\cite{taher2009accelerating}, which are now prevalent and widely available, could be game changing for accelerating BTE calculations of transport and nonequilibrium dynamics, and potentially a broader range of $e$-ph physics.
Unlike CPUs, which typically feature a limited number of high-performance cores and are optimized to handle complex workloads, GPUs can process a large volume of lightweight tasks. Originally designed for processing images, in the last decade GPUs have greatly expanded their scope in scientific computing, becoming an increasingly popular option for high-performance tasks~\cite{abramov2023november}. 
For example, on the Perlmutter cluster at the National Energy Research Scientific Computing Center (NERSC) \cite{nerscgpu}, at present about 40\% of the compute nodes are GPU nodes, each equipped with 28,000 CUDA cores. In contrast, each CPU node contains only 128 cores. Although each GPU core has lower computing power than a CPU core, GPUs can perform a large number of simple tasks with a high degree of parallelism. 
\\
\indent
However, two key considerations need to be addressed when designing algorithms for GPU execution. First, it is important to minimize data movement between the host and GPUs because it causes substantial overhead. Second, atomic operations must be optimized to avoid communication and synchronization between GPU cores, which causes significant performance loss~\cite{storti2015cuda}. 
This is particularly evident in a parallel CPU implementation of the BTE method, where different scattering channels contributing to the collision integral are typically handled by different processes or threads~\cite{zhou2021perturbo}.
Therefore, we seek to design a data structure for $e$-ph scattering in the BTE that is optimal for use on GPUs. 
\\
\indent
Several programming frameworks are available for GPU algorithms, such as the widely used CUDA and OpenACC. Due to its low-level architecture, CUDA offers greater control and optimization, but at the cost of increased complexity for code implementation and maintenance. In contrast, OpenACC offers a directive-based approach that enhances code readability and maintainability, with only a slight performance loss. 
In addition, OpenACC is designed for portability across platforms and thus is not limited to GPUs from any specific vendor. These strengths led us to use OpenACC in this work.
\\
\indent
Here, we design an efficient GPU data structure and algorithm for the BTE, and implement them with OpenACC in \textsc{Perturbo}, to accelerate calculations of transport and ultrafast dynamics using GPUs. Our new data structure optimizes data allocation and movement, as well as communication and synchronization between GPU cores. 
We show benchmarks for performance, memory consumption, and strong scaling in several materials. Our analysis shows a substantial performance improvement relative to the (already efficient) reference CPU implementation: we achieve a speed-up by $\sim$40 times for BTE calculations of transport and nonequilibrium dynamics on GPUs, realizing nearly linear scaling up to 100 GPUs. This new implementation was released in \textsc{Perturbo} v3.0 in early 2025. 
\\
\indent
\section{Results}
\subsection{BTE for transport and ultrafast dynamics}
\vspace{-3pt}
The semiclassical BTE models the change in electronic occupations in response to external fields and collision processes. In a solid, these processes are naturally described in momentum space~\cite{zhou2021perturbo}:
\begin{equation}\label{eq:bte1}
    \frac{\partial f_{n\boldsymbol{k}}(t)}{\partial t} = - \Big[ \hbar^{-1} \nabla_{\boldsymbol{k}}  f_{n\boldsymbol{k}}(t) \cdot \boldsymbol{F}\Big] + \mathcal{I}[f_{n\boldsymbol{k}}(t)],
\end{equation}
where $f_{n\boldsymbol{k}}(t)$ is the occupation factor at time $t$ of an electronic state with crystal momentum $\boldsymbol{k}$ and band index $n$~\cite{bernardi2016first,mahan2011condensed}. 
Note that we assume slowly varying fields and homogeneous material and electronic occupations, which removes spatial derivatives. Under an external field $\boldsymbol{F}$ ($\boldsymbol{F}= -e\boldsymbol{E}$ for electrons in the presence of an electric field $\boldsymbol{E}$), the change in electron occupations is determined by the drift term (first term on the right-hand side) and the collision integral $\mathcal{I}$. 
The BTE is solved in the time domain for ultrafast dynamics and at steady state for transport (see Methods for details). Computing the collision integral is the main bottleneck in the algorithm for both ultrafast dynamics, where it is computed at each time step, and for transport calculations, where it is computed in each iteration step.
\\
\indent
\subsection{Collision integral, \textit{e}-ph coupling matrix, and scattering channels}
\vspace{-3pt}
Using Fermi's golden rule, the collision integral $\mathcal{I}$ for nonequilibrium dynamics reads~\cite{mahan2011condensed, zhou2021perturbo}
\begin{equation}\label{eq:integraldyn}
    \begin{split}
        \mathcal{I}^{e\rm{-ph}}[f_{n\boldsymbol{k}}(t)] &= - \frac{2\pi}{\hbar} \frac{1}{\mathcal{N}_{\boldsymbol{q}}} \sum_{m\boldsymbol{q}\nu}  \Big|g_{mn\nu}(\boldsymbol{k},\boldsymbol{q})\Big|^2 \times \\
        \Big[
         \delta\Big(\varepsilon_{n\boldsymbol{k}} - & \varepsilon_{m\boldsymbol{k}+\boldsymbol{q}} + \hbar \omega_{\nu \boldsymbol{q}} \Big)\times F_{\rm abs}
         + \delta\Big(\varepsilon_{n\boldsymbol{k}} - 
        \varepsilon_{m\boldsymbol{k}+\boldsymbol{q}} - \hbar \omega_{\nu \boldsymbol{q}}\Big)\times F_{\rm em} \Big].
    \end{split}
\end{equation}

The above equation describes the $e$-ph scattering process for an electron from the initial state $\ket{n\boldsymbol{k}}$ with energy $\varepsilon_{n\boldsymbol{k}}$ to the final state $\ket{m\boldsymbol{k}+\boldsymbol{q}}$ with energy $\varepsilon_{m\boldsymbol{k}+\boldsymbol{q}}$, by absorbing (emitting) a phonon with wave vector $\boldsymbol{q}$ ($-\boldsymbol{q}$), mode index $\nu$, and frequency $\omega_{\nu\boldsymbol{q}}$. 
\\
\indent
Each absorption or emission process can be labeled using the notation $(\boldsymbol{k},\boldsymbol{q},n,m,\nu)$, here referred to as a scattering channel. 
The factors $F_{\rm{em}}$ and $F_{\rm{abs}}$ depend on the carrier occupations $ f_{n\boldsymbol{k}}(t)$ and the phonon occupations $N_{\nu\boldsymbol{q}}$. 
In addition, $g_{mn\nu}(\boldsymbol{k},\boldsymbol{q})$ are elements of the $e$-ph coupling matrix $\boldsymbol{g}(\boldsymbol{k},\boldsymbol{q})$ computed from first principles, the $\delta$ functions enforce energy conservation, and crystal momentum conservation is satisfied by using commensurate electron and phonon grids (with $\mathcal{N}_{\mathbf{q}}$ grid points) and selecting appropriate $(\boldsymbol{k},\boldsymbol{q})$ pairs for each scattering process. For convenience, in the following we denote the collision integral for state $\ket{n,\boldsymbol{k}}$ as $\mathcal{I}(n,\boldsymbol{k}) = \mathcal{I}^{ e\rm{-ph}}[f_{n\boldsymbol{k}}(t)]$.
Although the expression for $\mathcal{I}(n,\boldsymbol{k})$ is different in transport calculations (see Methods), the same data structure and algorithm apply to both transport and ultrafast dynamics, and are illustrated here for the ultrafast dynamics case.
\\
\indent
Computing the collision integral $\mathcal{I}$ for all electronic states involves summing over all active scattering channels. 
The electronic structure and lattice dynamics are first obtained in the entire Brillouin zone. Then, we restrict the electronic states of interest to a given energy window, significantly reducing the total number of $(\boldsymbol{k},\boldsymbol{q})$ pairs~\cite{zhou2021perturbo}.
For each $(\boldsymbol{k},\boldsymbol{q})$ pair, the nominal number of scattering channels prior to imposing any conservation constraints is $N_b^2 \times N_\nu$, where $N_b$ is the number of included bands and $N_\nu$ is the number of phonon modes. By imposing an approximate energy conservation (with a Gaussian $\delta$-function) and discarding channels with $\lvert g_{mn\nu}(\boldsymbol{k},\boldsymbol{q}) \rvert$ below a prescribed cutoff, the set of active scattering channels is further reduced to a small fraction of the total. This process makes $g_{nm\nu}(\boldsymbol{k},\boldsymbol{q})$ highly sparse in the parameter space $(\boldsymbol{k},\boldsymbol{q},n,m,\nu)$ (see Supplementary Fig. 1), saving extensive memory and computational cost~\cite{luo2024data}.
This scattering channel selection algorithm is implemented in \textsc{Perturbo} v2.2.0 and earlier CPU-based versions~\cite{zhou2021perturbo}, and achieves efficient performance and memory usage on CPUs. 
\\
\indent
However, the calculation of the collision integral $\mathcal{I}$ remains the most computationally demanding part of the BTE workflow and would greatly benefit from GPU acceleration. The CPU implementation is highly optimized using hybrid MPI and OpenMP parallelization but is not readily adapted to GPU parallelization for two main reasons.
First, when computing the collision integral, the CPU implementation uses a large number of atomic operations to avoid competition (so-called \lq\lq race conditions'') between threads parallelized over scattering channels.
Second, the use of numerous arrays of variable lengths to store information about the scattering channels $(\boldsymbol{k},\boldsymbol{q},n,m,\nu)$ requires referencing millions of individual heap allocations, introducing substantial overhead that limits GPU performance.
\\
\indent 
We propose a GPU-optimized data structure and algorithm that address these limitations and efficiently calculate the collision integral on GPUs. In the following, we describe this data structure and algorithm, and show benchmarks of performance, memory usage, and scaling behavior, using the CPU implementation of {\sc Perturbo} as a reference. 

\subsection{Data structure and implementation}
\subsubsection{Reference CPU algorithm}\label{sec:cuds}
\begin{figure}[ht!]
\includegraphics[width=1.0\textwidth]{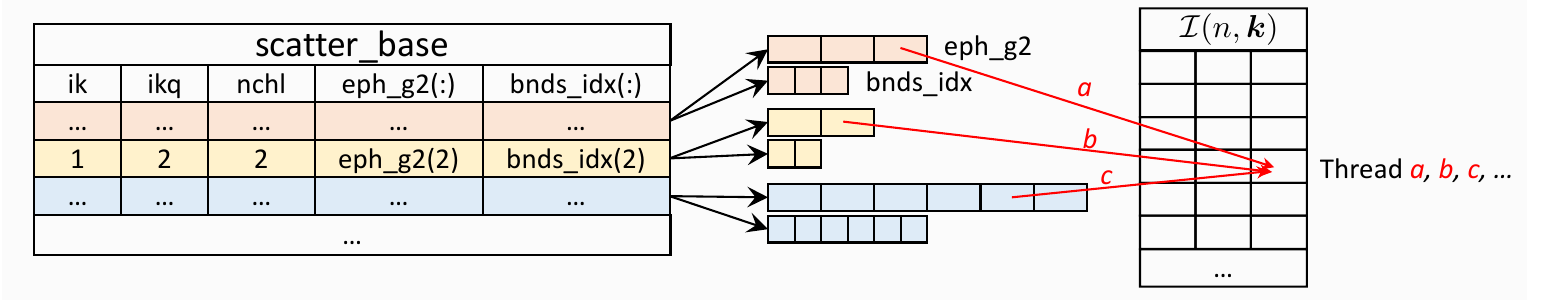}
\centering
\caption{\label{fig:dscpu} \textbf{Schematic of the \texttt{scatter\_base} data structure.} The information for each $(\boldsymbol{k}, \boldsymbol{q})$ pair is stored as a separate entry in \texttt{scatter\_base}, as shown using different colors. The variables \texttt{ik}, \texttt{ikq}, \texttt{nchl} are indices of the $\boldsymbol{k}$ and $\boldsymbol{k+q}$ points and the number of active scattering channels, respectively. The variables \texttt{eph\_g2} and \texttt{bnds\_idx} are arrays holding, respectively, the squared norm of the $e$-ph matrix elements and the joint indices of bands and phonon modes for each scattering channel. 
The relation of these variables to the collision integral $\mathcal{I}(n,\boldsymbol{k})$, whose components can be updated by multiple processes and threads simultaneously, is shown using red arrows.}
\end{figure}

\vspace{-8pt}
As discussed above, the electronic bands, phonon modes, and momenta collectively label an active scattering channel between states $\ket{n\boldsymbol{k}}$ and $\ket{m\boldsymbol{k}+\boldsymbol{q}}$.
The $e$-ph coupling matrix $\boldsymbol{g}$ is effectively sparse and irregular in this parameter space $(\boldsymbol{k},\boldsymbol{q},m,n,\nu)$. Consequently, using a single 5-dimensional array to store $g_{nm\nu}(\boldsymbol{k},\boldsymbol{q})$ leads to highly inefficient code as many of its entries are zero or very small (see Supplementary Fig.~1).
In practice, we group together all active scattering channels involving the same $(\boldsymbol{k}, \boldsymbol{q})$ pairs, and create an abstract type containing all relevant information for each pair. 
This design leverages benefits of object-oriented programming, such as flexibility and maintainability, while at the same time decoupling the $(\boldsymbol{k},\boldsymbol{q})$ pairs, which enables efficient distributed programming using MPI. 
Specifically, we define the object \texttt{scatter\_base} to store all the relevant information for all the active scattering channels of each $(\boldsymbol{k},\boldsymbol{q})$ pair, thus filtering out all redundant $e$-ph and scattering channel data.
A schematic of this data structure is shown in Fig.~\ref{fig:dscpu}.

We use this implementation and data structure, available in \textsc{Perturbo} v2.2.0 and earlier versions \cite{zhou2021perturbo}, as a reference or baseline for benchmarking code performance. This ``Baseline-CPU'' algorithm features hybrid MPI plus OpenMP CPU parallelization, where the $\boldsymbol{k}$ points are evenly distributed over different MPI processes. For ultrafast dynamics simulations, this code has two nested loops for each $\boldsymbol{k}$ point: an outer loop over $(\boldsymbol{k},\boldsymbol{q})$ pairs accelerated with OpenMP, and an inner loop over active scattering channels for the current $(\boldsymbol{k},\boldsymbol{q})$ pair, executed sequentially by each OpenMP thread. 
For transport, there is an additional inner loop over Cartesian components of the external field. When the same collision integral $\mathcal{I}(n,\boldsymbol{k})$ is updated by multiple threads simultaneously, as in the red arrows in Fig.~\ref{fig:dscpu}, the race condition between threads is avoided using OpenMP atomic operations. 

\begin{figure}[ht!]
\includegraphics[width=1.0\textwidth]{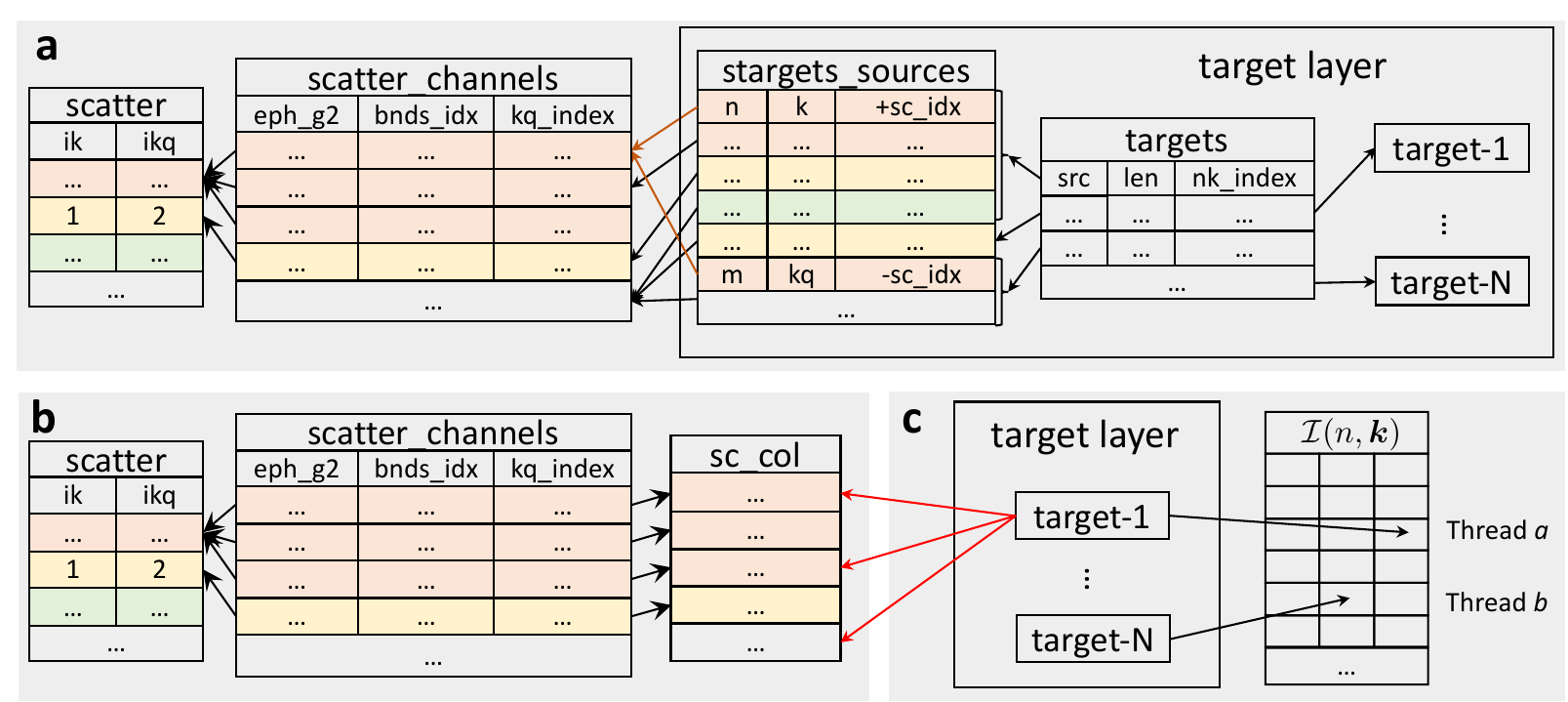}
\centering
\caption{\label{fig:dsoptgpu} \textbf{Data structure optimized for GPUs.} \textbf{a} Setup of the scattering channels and target layer. Relevant quantities for the $(\boldsymbol{k}, \boldsymbol{q})$ pairs are stored into multiple arrays: \texttt{scatter}, which stores the indexes of the $\boldsymbol{k}$ and $\boldsymbol{k+q}$ points, and \texttt{scatter\_channels}, which stores information for all scattering channels, such as the square of the $e$-ph matrix elements (\texttt{eph\_g2}), the joint indices of bands and phonon modes (\texttt{bnds\_idx}), and the index of the $(\boldsymbol{k}, \boldsymbol{q})$ pair (\texttt{kq\_index}). 
Scattering channels shown with the same color are associated with the same $(\boldsymbol{k}, \boldsymbol{q})$ pair. 
In addition, \texttt{stargets\_sources} indexes the elements of the collision integral $\mathcal{I}(n,\boldsymbol{k})$ and the position of the scattering channels (\texttt{sc\_idx}) in \texttt{scatter\_channels}. The positive (negative) sign of \texttt{sc\_idx} reflects how that entry contributes to the collision integral. 
The rows of \texttt{stargets\_sources} are arranged in order, with rows sharing the same $(n,\boldsymbol{k})$ grouped together, as shown with curly braces.
Each such group is called a target, and for each group, the position in \texttt{stargets\_sources} (\texttt{src}), the length (\texttt{len}), and the combined $(n,\boldsymbol{k})$ index (\texttt{nk\_index}) are stored in \texttt{targets}. Together, \texttt{stargets\_sources} and \texttt{targets} constitute the target layer. 
\textbf{b} Calculation of the contribution to the collision integral from each scattering channel, defined in Eq.~\ref{eq:integraldyn}, which is computed and stored in \texttt{sc\_col}. 
\textbf{c} Update of collision integrals $\mathcal{I}(n,\boldsymbol{k})$. Each element of $\mathcal{I}$ is updated by one target and one thread of execution. Using the target layer described in (\textbf{a}), each target is able to find the contribution of all the associated scattering channels in \texttt{sc\_col}, as shown with red arrows.}
\end{figure}

\subsubsection{Optimized GPU algorithm} \label{sec:nds}
A direct implementation of the Baseline-CPU algorithm on GPUs would be inherently inefficient as GPUs are optimized for executing many lightweight threads concurrently, and therefore perform poorly when frequent atomic updates are required. Moreover, \mbox{OpenACC} generates one data transfer per memory allocation, so allocating a large number of variable-length arrays incurs significant overhead. 
The combination of numerous heap allocations and frequent atomic operations across threads would severely limit the efficiency of a GPU version of the above algorithm.
\\
\indent
To achieve GPU acceleration, we redesign the data structure and code implementation for the key step of the BTE algorithm, the calculation of the collision integral. 
In the optimized data structure, shown in Fig.~\ref{fig:dsoptgpu}, we allocate \texttt{scatter} and \texttt{scatter\_channels}, which store the same information as \texttt{scatter\_base} but using fixed-size buffers instead of variable-length arrays. 
This avoids dynamic GPU allocations and improves performance, while preserving the flexibility of object-oriented programming. 
In addition, to eliminate atomic updates on the GPU, we develop an algorithm that inverts the accumulation scheme: rather than assigning multiple threads to update the contribution of each scattering channel to one collision integral $\mathcal{I}(n,\boldsymbol{k})$, it distributes threads over $\mathcal{I}(n,\boldsymbol{k})$ itself. Each thread then identifies the scattering channels that contribute to its assigned $\mathcal{I}(n,\boldsymbol{k})$. 
The calculation of the collision integral with the optimized GPU algorithm and data structure consists of three steps: 

\begin{enumerate} 
    \item \textit{Create the target layer:}
        \\
        In the first step, the scattering channels are traversed to determine to which elements of the collision integral $\mathcal{I}(n,\boldsymbol{k})$ they contribute. 
        Channels contributing to the same element of $\mathcal{I}$ are grouped together and form a \texttt{target}. 
        This grouping is handled in the target layer (Fig.~\ref{fig:dsoptgpu}a), where each scattering channel of \texttt{scatter\_channels} contributes to two different collision integrals, $\mathcal{I}(n,\boldsymbol{k})$ and $\mathcal{I}(m,\boldsymbol{k}+\boldsymbol{q})$. 
        These contributions are equal in magnitude but opposite in sign: In Fig. ~\ref{fig:dsoptgpu}a, the positive sign of \texttt{sc\_idx} denotes the contribution to $\mathcal{I}(n,\boldsymbol{k})$ and the negative sign to $\mathcal{I}(m,\boldsymbol{k}+\boldsymbol{q})$). 
        The rows of \texttt{stargets\_sources} are sorted based on their contribution to the collision integral and grouped into a \texttt{target}. 
        This step is not computationally intensive and is performed only once on CPUs. 
        See Supplementary Note 1 for demo code related to this step.

    \item \textit{Compute the contribution from each scattering channel:} 
    \\
    The second step evaluates the contribution of each scattering channel to the collision integral. 
    As shown in Fig.~\ref{fig:dsoptgpu}b, the contribution to $\mathcal{I}(n,\boldsymbol{k})$ from each channel is stored in the variable \texttt{sc\_col}, which has the same number of rows as \texttt{scatter\_channels}. This computationally demanding step is accelerated on GPUs using heterogeneous programming with OpenACC. Example code is provided in Supplementary Note 2.

    \item \textit{Collect the contributions from all targets:}
    \\
    This step updates the collision integrals using contributions from all targets computed in Step 2. The routine loops over \texttt{targets} to update the elements of $\mathcal{I}$ by summing over contributions from all scattering channels in each target (Fig.~\ref{fig:dsoptgpu}c). 
    This way, only one element of $\mathcal{I}$ will be updated for each thread, as shown with black arrows in Fig. \ref{fig:dsoptgpu}c. Using \texttt{target}-1 as an example, the first row of \texttt{targets} records the position of this target in \texttt{stargets\_sources}. 
    By using the value of \texttt{sc\_idx} for all relevant elements in \texttt{stargets\_sources}, the code finds all the scattering channels in \texttt{sc\_col} (see red arrows in Fig.~\ref{fig:dsoptgpu}c). Then those values are summed together to update the corresponding element of $\mathcal{I}$ in the current thread. This step is performed on GPUs for acceleration. 
    Demo code for this step is provided in Supplementary Note 3.
\end{enumerate}

\indent
The new data structure resolves the inefficiency of the Baseline-CPU method by minimizing host–device data transfers, reducing the number of atomic operations across threads, as well as eliminating memory padding through data alignment.
\\
\indent
As our discussion has focused on ultrafast dynamics, we briefly mention the main differences in the implementation for transport calculations. For ultrafast dynamics, the contribution to the collision integrals $\mathcal{I}(n,\boldsymbol{k})$ and $\mathcal{I}(m,\boldsymbol{k}+\boldsymbol{q})$ from a single scattering channel is equal in magnitude and opposite in sign. 
This allows us to define only a 1D array for \texttt{sc\_col}(:) and use the sign of \texttt{sc\_idx} for bookkeeping. 
For transport, this is not possible, and thus \texttt{sc\_col} needs an additional dimension. In practice, \texttt{sc\_col} is defined as a 3D array to account for the external field, and the code has an additional loop over the directions of the field. These small differences do not affect the performance.\\

\begin{figure}[t]
\includegraphics[width=1.0\textwidth]{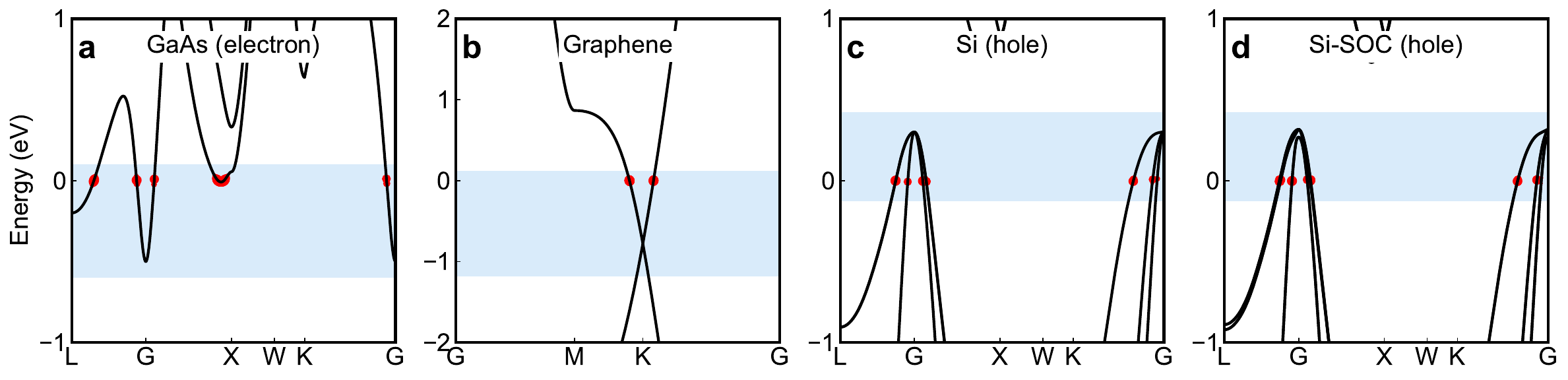}
\centering
\caption{\label{fig:material}\textbf{Simulation setup for four systems.} \textbf{a} Electrons in GaAs, \textbf{b} electrons in graphene, \textbf{c} hole carriers in silicon, and \textbf{d} holes in silicon with SOC. Band structures are shown together with the selected energy windows (shaded regions) and the initial populations for the nonequilibrium dynamics simulations (red dots). Energies are shifted so that the Fermi energy is at 0 eV.}
\end{figure}

\subsection{Performance, memory usage, and scaling analysis of GPU code}\label{sec:results}
\subsubsection{Simulation setup}
\vspace{-6pt}
We benchmark the GPU implementation on four selected systems: electron carriers in gallium arsenide (GaAs), graphene, hole carriers in silicon (Si) modeled without spin-orbit coupling (SOC), and hole carriers in silicon with SOC (Si-SOC). These cases cover a range of scenarios, including metals and semiconductors, calculations with and without SOC, electron and hole carriers, and 2D and bulk materials. The band structures, energy windows for nonequilibrium dynamics, and the initial population for nonequilibrium simulation for all four systems are shown in Fig.~\ref{fig:material}. 
\\
\indent
As shown in Table~\ref{tab:gs}, in the ultrafast dynamics simulation of electrons in GaAs, after imposing an energy window of 0.7 eV above the conduction band edge, the number of $\boldsymbol{k}$ and $\boldsymbol{q}$ points in GaAs are reduced from $135^3$ to 56713 (2\% of the original value) and 637412 (26\% of the original value), respectively. Imposing energy conservation and a cutoff on $|\mathbf{g}(\boldsymbol{k},\boldsymbol{q})|$ further reduces the number of scattering channels, from a nominal value of $10^{11}$ to an actual value of $10^8$. These values justify our approach of keeping only the active scattering channels ($10^8$ in the case of GaAs) and looping over these channels in the code.

\begin{table}[ht]
\centering
\caption{\label{tab:gs}Summary of simulation parameters for the four systems studied, including the number of bands and phonon modes (n,$\nu$), the number of $(\boldsymbol{k},\boldsymbol{q})$ points, and the number of nominal and active scattering channels. \vspace{6pt}}
\setlength{\tabcolsep}{6pt} 
\begin{tabular}{c c c c c c} 
 \hline\hline
 \multirow{2}{*}{Systems} &\multirow{2}{*}{(n,$\nu$)}  & \multicolumn{2}{c}{$(\#\boldsymbol{k},\#\boldsymbol{q})$} & \multicolumn{2}{c}{\# channels} \\
&& Nominal & Energy window &Nominal& Active\\
 \hline
GaAs &(1,6)& ($135^3$, $135^3$) &(56713, 637412) &$10^{11}$& $10^{8}$\\
 \hline
graphene &(1,6)&($1300^2$, $1300^2$)  &(43206, 131605)& $10^{11}$ &$10^{8}$\\
\hline
Si &(3,6)&($105^3$, $105^3$) &(44275, 232728) &$10^{12}$&$10^{8}$\\
\hline
Si-SOC &(6,6)&($95^3$, $95^3$) &(29317, 151560)&$10^{12}$&$10^{7}$ \\
\hline
\hline
\end{tabular}
\end{table}

\subsubsection{Performance comparisons}
\begin{figure}[t!]
\includegraphics[width=1.0\textwidth]{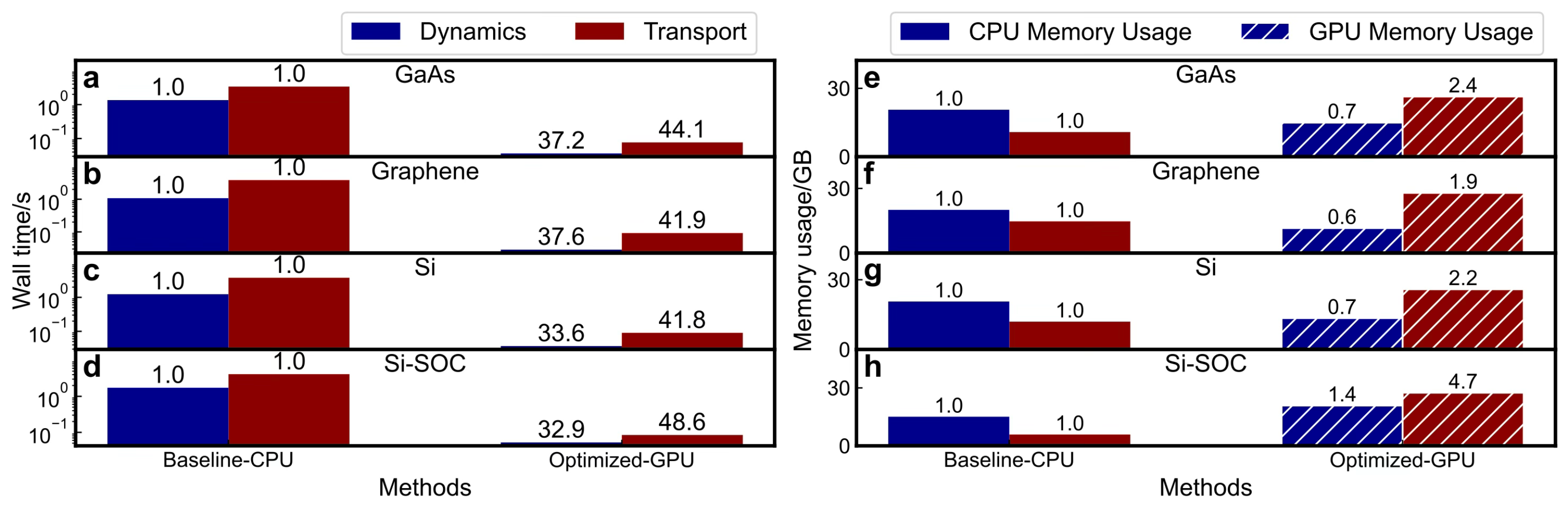}
\centering
\caption{\label{fig:perform}
\textbf{Performance of the optimized GPU implementation.} \textbf{a}-\textbf{d}, performance, and \textbf{e}-\textbf{h}, memory usage, for the four systems studied here, respectively. 
The left panels, \textbf{a}–\textbf{d}, show the wall time (in seconds, on a logarithmic scale) for ultrafast dynamics (blue) and transport (red) calculations. The speedup values, obtained as the ratio of Baseline-CPU to optimized-GPU code wall times, are given above each bar in the optimized-GPU results.
The right panels, \textbf{e}-\textbf{h}, give the memory usage (in GB) on CPU (solid colors) and GPU (striped bars) for the same systems. Memory usage values annotated in the plot are referenced to the baseline CPU results.}
\end{figure}

We first compare the wall-time of the optimized-GPU code with the Baseline-CPU algorithm to directly show the performance improvement. For a fair comparison, both the CPU and GPU calculations are conducted on a single heterogeneous GPU node using the same hardware (one node includes four A100 GPU chips and one AMD CPU with 64 cores). See Methods for details. 
Figure~\ref{fig:perform}a-d compare calculations carried out with the baseline CPU algorithm, used as a reference, and the optimized GPU algorithm, for both transport and ultrafast dynamics, for the four systems studied.
For transport calculations, the wall time used in the plots is the average elapsed wall time of each iteration in the iterative BTE solution, and for ultrafast dynamics, the wall time is for one time step of the rt-BTE simulation.\vspace{10pt}

The speed up of the GPU implementation relative to the baseline CPU code is noteworthy. Our optimized GPU code achieves a speedup by a factor of 44 for transport and 35 for ultrafast dynamics. 
We find similar speed-ups for all systems in our test set, showing that the speed-up is an intrinsic feature of the GPU algorithm independent of the system studied. This order-of-magnitude speed up is the result of careful design and optimization of array structures, data transfer, and thread management in our GPU algorithm.
\\
\indent
To demonstrate the importance of our novel data structure optimized for GPUs, we run the optimized GPU code with and without OpenACC directives in the same HPC settings. 
This test compares the performance of the same GPU-optimized code executed on GPU versus CPU hardware; this is a fair and established approach to compare CPU and GPU code. As shown in the Supplementary Fig. 2, the optimized-GPU code runs 25$-$50 times faster on GPUs than on CPU hardware, for both transport and ultrafast dynamics. This result demonstrates the considerable acceleration achieved on GPUs. 
\\
\indent
Next, we compare memory consumption in the baseline CPU and optimized GPU algorithms. In Fig.~\ref{fig:perform}e-h, we show data on memory usage. 
For all calculations without SOC, we find that the memory allocation in the optimized GPU code is approximately 70\% for ultrafast dynamics, and 200\% for transport, relative to the memory used in the baseline CPU code. 
This difference arises from the need to store several intermediate variables in the GPU implementation of transport, in particular the direction of the applied electric field, which in the Baseline-CPU code is replaced by an iterative loop without loss of performance. Finally, the calculation with SOC in Fig.~\ref{fig:perform}h uses more memory than the cases without SOC. 
The reason is that the number of scattering channels in the arrays scales with the number of bands, resulting in a higher memory usage when SOC is included. Despite the higher memory usage, the wall-time speed-up is unchanged in the presence of SOC.

\subsubsection{Strong scaling analysis}
\vspace{-10pt}
The strong scaling measures how the speedup scales with the number of computing nodes for a fixed simulation size. Therefore, strong-scaling benchmarks address a key question for computationally intensive simulations: What performance improvement can one obtain by increasing the computational resources? 
To avoid confusion with the speed-up of GPU versus CPU code discussed above, we define the strong-scaling speed-up as: 

\begin{equation}
    \text{Strong-scaling speedup}_N = \frac{T_{\rm 4-nodes}}{T_{\rm N-nodes}}\vspace{12pt}
\end{equation}
where $T_{\rm 4-nodes}$ and $T_{\rm N-nodes}$ are the wall times for simulations using 4 and $N$ nodes, \mbox{respectively.} 
\begin{figure}[t]
\includegraphics[width=1.0\textwidth]{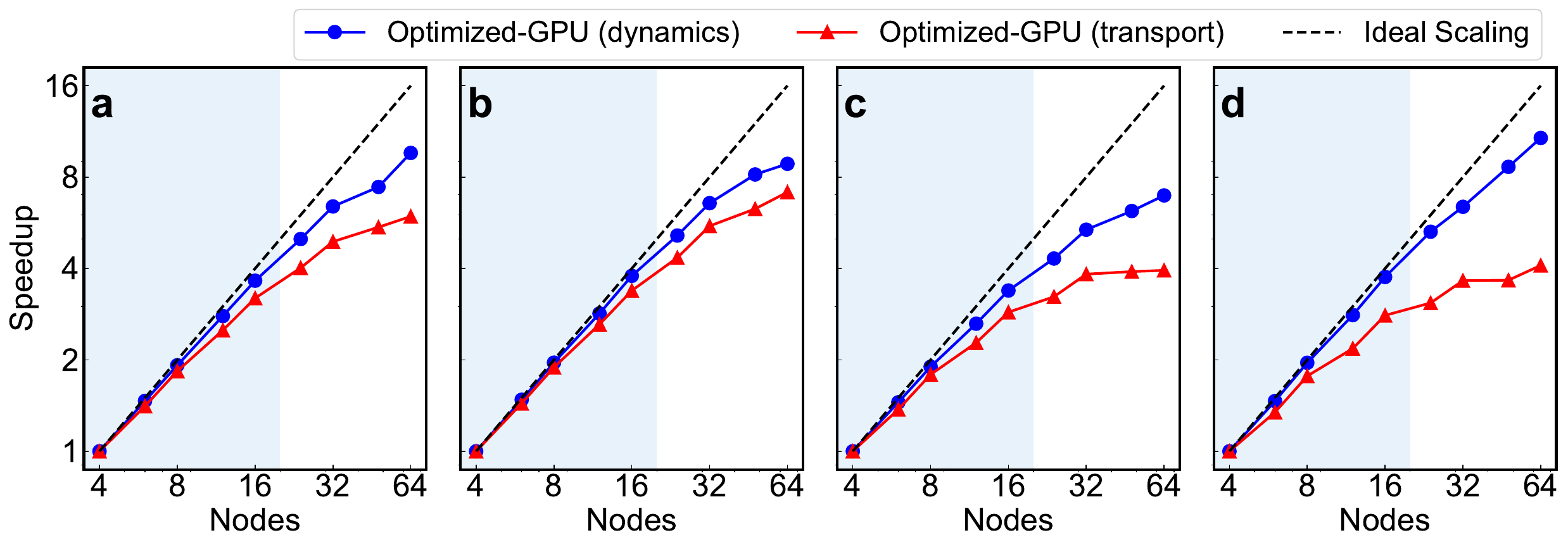}
\centering
\caption{\label{fig:ssa} \textbf{Strong-scaling performance.} Speedup versus number of GPU nodes for \textbf{a} GaAs, \textbf{b} graphene, \textbf{c} Si, and \textbf{d} Si with SOC. Results for the optimized-GPU code are shown using solid lines with symbols for ultrafast dynamics (purple) and transport (red). The dashed line shows the ideal linear scaling. Common scenarios for most users ($\leq 20$ GPU nodes) are indicated with shaded regions.}
\end{figure}

We carry out strong-scaling benchmarks for the optimized GPU code, for both transport and ultrafast dynamics, using between 4 and 64 GPU nodes. Setup details of these simulations, which employ very dense grids in momentum space, are provided in Table \ref{tab:perf1n}. 
Strong scaling results for the four systems studied here are shown in Fig.~\ref{fig:ssa}. 
Based on the definition of strong-scaling speedup given above, the ideal speedup is $\frac{N}{4}$, where $N$ is the number of nodes. This ideal value is shown in Fig.~\ref{fig:ssa} with a dashed line and used as a reference. 
\\
\indent
Our results in Fig.~\ref{fig:ssa} show that in common scenarios for most users, consisting of calculations with up to 20 GPU nodes, our GPU code exhibits nearly ideal scaling performance, which extends up to 24 nodes or more nodes in most cases. The only case that deviates from this trend is transport calculations in Si with or without SOC (red lines in Fig.~\ref{fig:ssa}c-d). Due to the high memory demand for these calculations, the scaling remains nearly ideal up to 8 GPU nodes, but deviates increasingly beyond that. 
For all systems, the acceleration efficiency drops to around 40–60\% of the ideal scaling at 64 nodes. This reduction is common in GPU codes and is mainly due to insufficient workload per GPU when such a large number of nodes is employed. 
This result implies that allocating excessive GPU resources to a calculation that can be completed on a much smaller number of nodes is unnecessary and inefficient.\vspace{24pt} 

\begin{table}[b]
    \centering
    \caption{\label{tab:perf1n}Momentum grid size and computational resources for performance and strong scaling analysis on four systems. The grid size for each material is chosen to maximize memory usage on one node for performance analysis. Note that the momentum grid is two-dimensional in graphene and three-dimensional for the other materials.\vspace{8pt}}
    \setlength{\tabcolsep}{6pt} 
    \begin{tabular}{c c c c c c} 
     \hline\hline
    
    & & GaAs & graphene&Si&Si-SOC \\
     \hline
    Performance& Transport &120$^3$&1200$^2$&95$^3$&75$^3$ \\
     \cline{2-6}
    (1 node, size fixed)& Dynamics &135$^3$&1300$^2$&105$^3$&90$^3$\\
     \hline
    Strong scaling& Transport &155$^3$&1700$^2$&125$^3$&95$^3$ \\
     \cline{2-6}
    (\#nodes vary, size fixed)& Dynamics &195$^3$&2300$^2$&150$^3$&120$^3$\\
    \hline
    \hline
    \end{tabular}
\end{table}

\clearpage
\section{Discussion}
In this work, we have developed an efficient GPU algorithm and data structure to accelerate BTE calculations of electronic transport and ultrafast dynamics governed by $e$-ph interactions. The code is developed in OpenACC and is included in version 3.0 of the open-source code \textsc{Perturbo}. Due to the directive-based approach used in OpenACC, the implementation is easy to maintain. With GPU acceleration, this new version of \textsc{Perturbo} provides a highly efficient MPI+OpenMP+GPU parallelization that can fully take advantage of modern Exascale HPC computing environments with multi-core CPUs and GPU accelerators.
\\
\indent
Through extensive benchmarks, we report speed-ups by
a factor of $\sim$40 for transport and ultrafast dynamics relative to the reference, state-of-the-art CPU implementation in the previous version of {\sc Perturbo}. 
This result is achieved by reformulating the data structure and algorithm to store $e$-ph interactions and carry out calculations of collision integrals. 
Our approach optimizes data allocation and movement between host and GPUs and synchronization between numerous GPU cores. We analyze the performance, memory consumption, and strong scaling for three materials. The strong scaling analysis shows nearly ideal scaling up to 16 GPU nodes (64 GPUs), with only a slight decrease in performance up to 24 GPU nodes (96 GPUs) for most cases. 
\\
\indent
While the optimized GPU data structure is discussed in the context of $e$-ph interactions, the same data structure can also be used for other scattering mechanisms, such as electron-defect~\cite{lu2019efficient}, exciton-phonon \cite{chen2020exciton}, magnon-phonon \cite{le2025magnon}, and phonon-phonon \cite{tong2021toward,kelly-ph} interactions. Note that we did not consider GPU acceleration for the interpolation of $e$-ph matrices, which has been studied in previous work~\cite{cepellotti2022phoebe, liu2023high}, mainly because recently developed compression algorithms~\cite{luo2024data} make the $e$-ph interpolation routines already highly efficient. 
In a future release, we will extend the GPU acceleration feature to other modules of \textsc{Perturbo}. 
\clearpage

\vspace{-6pt}
\begin{acknowledgements}
This work was supported by the National Science Foundation under Grant No. OAC-2209262. 
The ultrafast carrier dynamics calculations are based on work performed within the Liquid Sunlight Alliance, which is supported by the U.S. Department of Energy, Office of Science, Office of Basic Energy Sciences, and Fuels from Sunlight Hub under Award DE-SC0021266. 
This research used resources of the National Energy Research Scientific Computing Center, a DOE Office of Science User Facility supported by the Office of Science of the U.S. Department of Energy under Contract No. DE-AC02-05CH11231 using NERSC award NERSC DDR-ERCAP0026831.

\section{Competing interests}
The authors declare no competing interests.

\end{acknowledgements}

\setcounter{equation}{0}
\setcounter{figure}{0}
\setcounter{table}{0}

\end{document}